\documentclass[twocolumn,prb,aps,showpacs,superscriptaddress]{revtex4}
\usepackage{graphicx}

\begin{document}

\title{Penetration of Andreev bound states into the ferromagnet in a
    SrRuO$_{3}$/(110)YBa$_2$Cu$_3$O$_{7-\delta}$ bilayer: a scanning
    tunneling spectroscopy study
}

\author{Itay Asulin}
\affiliation{Racah Institute of Physics, The Hebrew University,
Jerusalem 91904, Israel}

\author{Ofer Yuli}
\affiliation{Racah Institute of Physics, The Hebrew University,
Jerusalem 91904, Israel}

\author{Israel Felner}
\affiliation{Racah Institute of Physics, The Hebrew University,
Jerusalem 91904, Israel}

\author{Gad Koren}
\affiliation{Department of Physics, Technion - Israel Institute of
Technology, Haifa 32000, Israel}

\author{Oded Millo}
\email{milode@vms.huji.ac.il} \affiliation{Racah Institute of
Physics, The Hebrew University, Jerusalem 91904, Israel}

\begin{abstract}
Scanning tunneling spectroscopy of thin epitaxial
$SrRuO_{3}/(110)YBa_2Cu_3O_{7-\delta}$ ferromagnet/superconductor
bilayers, reveal a clear penetration of the Andreev bound states
into the ferromagnetic layer. The penetration is manifested in the
density of states of the ferromagnet as a split zero bias
conductance peak with an imbalance between peak heights. Our data
indicate that the splitting occurs at the superconductor side as a
consequence of induced magnetization, confirming recent
theoretical predictions. The imbalance is attributed to the spin
polarization in the ferromagnet.
\end{abstract}


\maketitle
\section{Introduction}
The study of superconductor (S)/ ferromagnet (F)
proximity systems allows a direct investigation of the interplay
between the two competing orders of superconductivity and
ferromagnetism. In a highly transparent S/normal-metal (N)
junction, superconducting correlations are induced in N while they
are weakened in the S side. The mechanism underlying the proximity
effect (PE) is the Andreev reflection (AR): a hole-like
quasiparticle impinging on the interface from the N side is
retro-reflected as an electron-like quasiparticle with opposite
momentum and spin while destroying a Cooper pair in the S.
Consequently, in S/F junctions, the properties of the PE are
significantly modified due to the spin polarization and the
presence of an exchange field in F. Theoretical works predict a
rapid oscillatory decay of the induced superconducting order
parameter inside F. \cite{Demler,Buzdin, Zareyan} These
predictions, that were confirmed by various experiments
\cite{Kontos,Ryazanov,Freamat}, stem from the singlet pairing in
the S, and hold for both \textit{s}-wave and \textit{d}-wave
(along antinodal directions) superconductors.\cite{Stefanakis,
Faraii} However, the understanding of how the anisotropy of the
\textit{d}-wave symmetry within the a-b plane manifests itself in
the PE in the presence of an exchange field, is still rudimentary.
In particular, it is still unclear what happens to the Andreev
bound states (ABS) that reside at the nodal surfaces of
\textit{d}-wave S and how, in turn, they affect the density of states (DOS) of F.\\
\indent Splitting the current carrying ABS requires the removal of
either their directional or spin degeneracy. The former may result
from the admixture of a subdominant order parameter \cite{Amos2}
while the latter may be due to a small magnetization at the
surface. A possible origin of such a surface magnetization is the
background antiferromagnetic correlations in the underdoped regime
of the high-$T_{c}$ superconductors. \cite{Honerkamp}
Alternatively, a magnetization inside the S may also result from a
proximity to a magnetic layer. Recent theoretical works that
considered the inverse PE in S/F bilayers predict an induced
magnetization in the S side over a length-scale of the
superconducting coherence length, $\xi_{S}$.
\cite{Kharitonov,Bergeret1,Bergeret2,Krivoruchko} The direction of
the induced magnetic moment near the interface may be parallel or
antiparallel to the magnetization in F, depending on the interface
transparency, the thickness of F and the strength of the exchange
field. Experimentally, there is little evidence for such an
induced magnetization.\cite{Stahn,Stamopoulos} The induced
magnetization is expected to remove the spin degeneracy of the ABS
at the S/F interface and shift them to finite energies. However,
this effect was not observed so far in the DOS of F and is the
focus of this paper.\\
\indent We have previously shown in N/nodal-oriented
\textit{d}-wave S bilayers that as long as the interface is highly
transparent, the N layer is ballistic and phase coherence is
maintained, ABS can penetrate into the N layer and give rise to a
zero-bias conductance peak (ZBCP) in its DOS. \cite{Asulin} The
question that arises now is what happens when an exchange field is
introduced in the normal layer. The possible penetration of the
(spin degenerate) ABS into a spin polarized F layer was not
observed experimentally and remains unclear theoretically.\\
\indent In this work we employed scanning tunneling spectroscopy
on thin epitaxial $SrRuO_{3}/(110)YBa_2Cu_3O_{7-\delta}$ \
(SRO/(110)YBCO) S/F bilayers below full coverage of the YBCO by
the SRO layer. Our spectroscopy measurements clearly reveal an
intricate penetration of the ABS into the F layer to distances of
at least 9 nm. The penetration is reflected in the DOS of the F
layer as a split ZBCP with a pronounced imbalance between split
peak heights. The split is attributed to the removal of the spin
degeneracy of  the ABS due to induced magnetization in the S. The
imbalance between the peak heights is attributed to the spin
polarization in the F layer.
\begin{figure}[t]
\includegraphics[width=8.5cm]{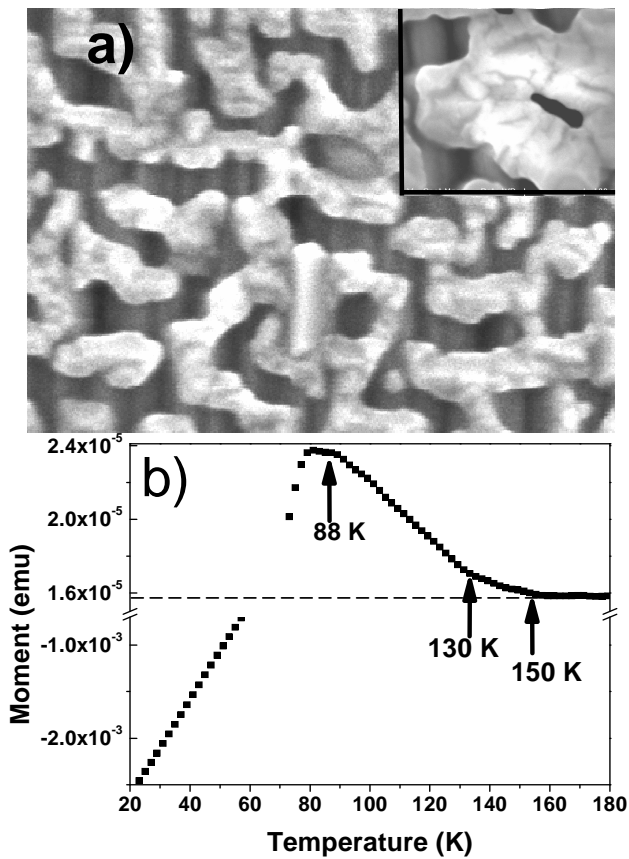}
\caption{(a) SEM image (0.8X0.6 $\mu m^2$) of a 3 nm thick SRO
layer on a (110)YBCO film demonstrating the SRO island (bright
areas) topography. Inset: a zoom on one SRO crystallite. (b)
Magnetization measurement (at 1000 Oe after zero-field cooling)
depicting the ferromagnetic transition onset at 150 K and the S
transition at 88 K.}\label{fig1}
\end{figure}
\section{Experiment}
\indent SRO is an itinerant ferromagnet ($ T_{Curie}=150 K$ in
thin films) with lattice parameters that are similar to those of
YBCO \cite{Zakharov} and therefore they can form highly
transparent interfaces, essential for the existence of AR and the
PE. A total of 16 bilayers of SRO (2-17 nm nominal thickness) on
(110) YBCO were prepared by laser ablation deposition on (110)
$SrTiO_3$ substrates. In order to achieve YBCO layers with the
nodal (110) orientation, first a 10 nm thick template layer of
YBCO was deposited at 660$^\circ$C, and then another YBCO layer of
50 nm thickness was deposited at 750$^\circ$C. The (110)
orientation was verified by x-ray diffraction. Finally, the SRO
layer was deposited at 785$^\circ$C. Annealing for obtaining
optimally doped YBCO was done \textit{in-situ} under 50 Torr
oxygen gas pressure and a dwell of one hour at 450$^\circ$C (the
SRO layer is insensitive to the oxygen annealing process). The
bilayers showed S transition temperatures at around 88 K with a
transition width of about 3 K, implying nearly optimally doped
films. The tunneling spectra (dI/dV vs. V characteristics,
proportional to the local quasi-particle DOS) were obtained at 4.2
K, much lower than both the superconducting and ferromagnetic
transitions.
\section{RESULTS AND DISCUSSION}
\indent The bare (110)YBCO films feature $\sim40\times100$
nm$^{2}$ elongated crystallite with uniform directionality over
areas of a few $\mu$m$^{2}$ (see Ref. 17). Figure 1a presents a
scanning electron microscope (SEM) image of a (110)YBCO film
covered with a 3 nm thick (nominal) SRO layer. At low thicknesses
the SRO layer exhibits island-film topography where the underlying
elongated crystallite structure of the (110)YBCO is clearly
apparent in between the islands. This island topography enabled us
to measure, on the same sample, the evolution of the DOS from that
of a bare (110)YBCO surface to the DOS of a ferromagnet in
proximity to a (110) surface. Full coverage by the SRO layer was
obtained at average thicknesses of above 10 nm. Below full
coverage, the islands featured varying thickness even within each
island. This is apparent in the inset of Fig. 1b which focuses on
one SRO crystallite. It is important to note that in thin SRO
films of thicknesses lower than 5 unit cells, $T_{Curie}$
decreases substantially.\cite{Izumi} Therefore it is expected that
the exchange field and local magnetization in films that feature
thickness variations at these ranges will not be uniform. Fig. 1b
depicts a SQUID magnetometer magnetization measurement performed
on a 3 nm thick SRO layer over-coating a (110)YBCO layer. The
onset of the ferromagnetic transition at $\sim150$ K is clearly
observed, indicating that our SRO layers at these thicknesses are
ferromagnetic. However, it is also apparent that there is a
distribution of the value of $T_{Curie}$: the transition is not
sharp and takes place over a range of temperatures starting from
150 K down to 130 K (marked by arrows in Fig. 1b), indicating that
the SRO layer is magnetically inhomogeneous. This inhomogeneity
stems from two main factors. First is the island topography that
yields differences between the mean exchange field at the
crystallite edge and its interior. Second is the thickness
variations, as discussed above. \cite{Izumi}
The diamagnetism of the YBCO sets in at about 88 K. \\
\indent The tunneling spectra of the bare (110)YBCO films featured
a pronounced ZBCP with gap-like features all over the YBCO
crystallites (see Ref. 17 for data on similar films).
\begin{figure}[t]
\includegraphics[width=8.5cm]{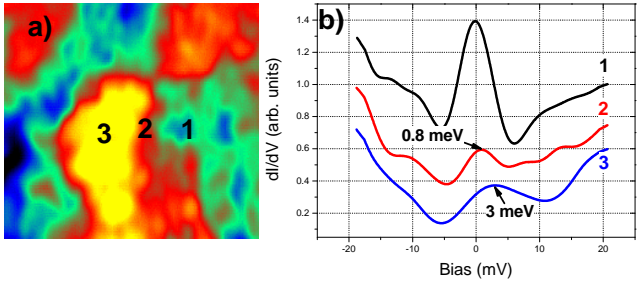}
\caption{(color online) (a) 80 $nm^2$ STM image of a 2 nm thick
SRO layer on a (110)YBCO film showing an SRO crystallite (yellow
brighter area). (b) Tunneling spectra (vertically shifted for
clarity) obtained at the points marked in (a)}\label{top}
\end{figure}
Interestingly, ABS were also detected in the tunneling spectra
obtained on the SRO crystallites. Figure 2a presents an STM image
of a 2 nm thick SRO layer overcoating a (110)YBCO film where an
SRO crystallite is apparent. The tunneling spectra presented in
Fig. 2b were taken at the points marked in the topographic image.
Far enough from the crystallite (point 1), on a bare YBCO region,
a pronounced centered ZBCP was measured (upper black curve),
similar to those measured on bare (110) films.\cite{Asulin} The
middle (red) dI/dV curve was obtained at the vicinity of the
crystallite edge where the thickness of the SRO is lower compared
to the center of the crystallite (point 2). The amplitude of the
peak is largely suppressed and it is shifted by $\sim0.8$ meV to
positive energies. The lower (blue) dI/dV curve was obtained on
the SRO crystallite (point 3). Here, the ZBCP is shifted to +3 meV
and the peak appears to be broadened. In fact, our overall
accumulated data (see below) implies that this shifted and
broadened peak reflects a split ZBCP in which the negative energy
peak (in this case) is largely suppressed and smeared (due to noise and life-time broadening effects). \\
\indent The typical tunneling spectra that were obtained on the
SRO crystallites featured a pronounced split of the ZBCPs with a
varying degree of imbalance between the positive and negative peak
heights. This imbalance ranges from a nearly symmetrical,
double-peak structure to an asymmetrical split in which one of the
peaks is fully suppressed. Interestingly, the suppression of both
the positive and negative peaks were observed with roughly the
same occurrence (even on the same sample). Moreover, the amplitude
of the split (energy shift of ABS) also featured pronounced
changes ranging between less than 1 meV and a maximum energy shift
of $\sim 5$ meV (much larger than the known split amplitude due to
the admixture of a sub-dominant OP at optimal doping).\cite{Amos2}
We note that a wide range of both the degree of imbalance and the
energy shift were observed on all samples with average SRO
thickness up to 9 nm. Figure 3 shows the range of peak imbalance
and the suppression of negative as well as positive peaks. Here,
the dI/dV curves were all obtained on the same sample (but in
different areas), with an SRO thickness of 3 nm. The peaks in Fig.
3a (with shifts of up to $\sim3.5$ meV) exhibit a small, but
pronounced, imbalance between peak heights where the positive peak
is suppressed relative to the negative peak. The imbalance is
larger in Fig. 3b in which the positive energy peak seems to be
fully suppressed and the negative peak is found at -3.1 meV. The
degree of imbalance in Fig. 3c is roughly the same as in 3a but
here the negative energy peak is suppressed. In Fig. 3d the
suppression is, again, at the negative side and only a trace of
the negative peak is still detectable, yet more than in 3b.\\
\begin{figure}[]
\includegraphics[width=8.5cm]{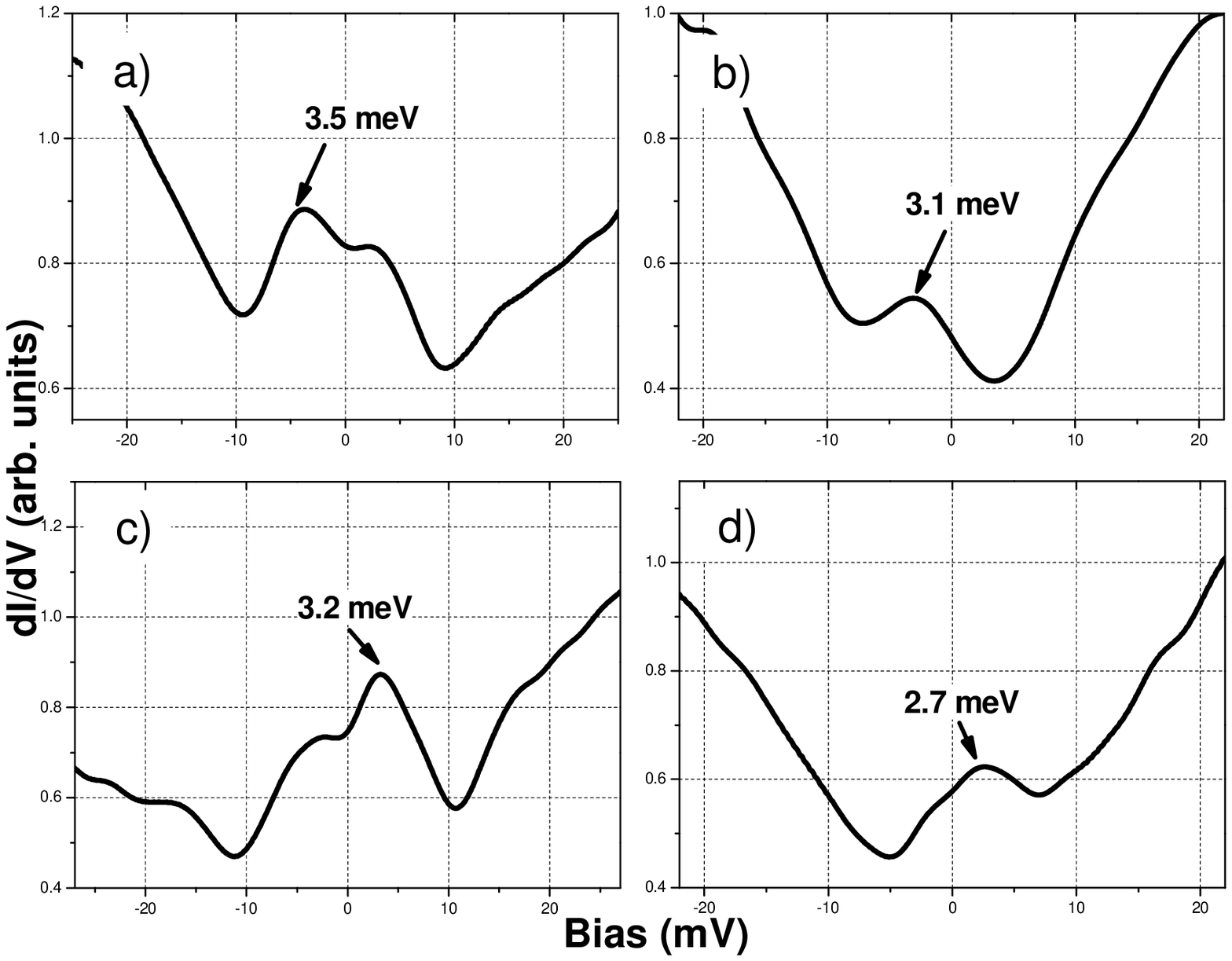}
\caption{ (color online) Tunneling spectra obtained on a 3 nm
thick SRO layer overcoating a (110)YBCO film demonstrating the
range of imbalance between split ZBCP peak heights and the
suppression of both positive and negative peaks.}\label{splits}
\end{figure}
\indent The range of the energy shift is demonstrated in Fig. 4.
Here, the dI/dV curves were acquired sequentially at equal steps
along a 200 nm long line taken on a 2 nm thick SRO layer. The ZBCP
evolves from a centered ZBCP (bottom curve) into a shifted ZBCP
where the energy shift increases up to a value of 4.5 meV (upper
curve). This variation of the energy shift occurs on a single SRO
island (as demonstrated also in Fig. 2). It is important to note
that there is no clear correlation between the magnitude of the
split (energy shift) and the degree of imbalance between peak
heights (e. g., compare figure 3a and 3b),
suggesting that these two effects have a different origin. \\
\begin{figure}[]
\includegraphics[width=8cm]{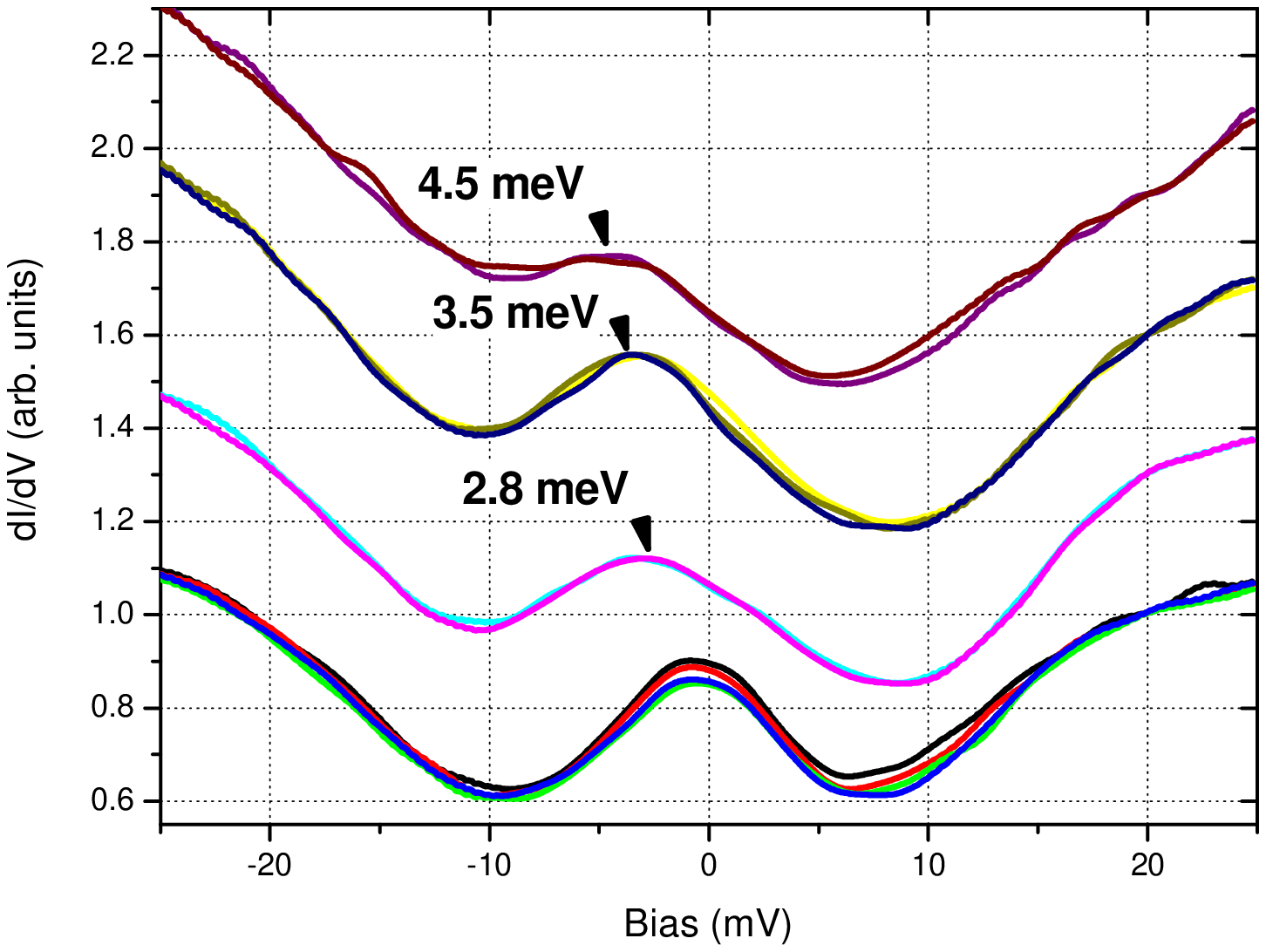}
\caption{Tunneling spectra obtained sequentially at fixed steps
along a 200 nm line on a 2 nm thick SRO layer overcoating a
(110)YBCO film demonstrating the range of energy shift. The curves
are shifted vertically for clarity.}\label{continuous}
\end{figure}
\indent We shall first address the mechanism by which ABS can
penetrate the F layer. The formation of ABS at the bare nodal
(110) surface in YBCO can be understood within the framework of a
quasiclassical insulator (I)/N/S quantum well model\cite{Hu},
where an effective N layer of thickness $\xi_{S}$ results from the
pair breaking nature of nodal surfaces. Here, the current carrying
ABS correspond to closed quasiparticle trajectories consisting of
ARs at the S/N interface and normal reflections at the free
interface. A metallic N layer, that is in good electrical contact
with the S, serves as an extension of the above-mentioned inherent
N region at the nodal surface and the INS model is still valid.
\cite{Asulin} The picture is more intricate if the N layer is
replaced with a spin polarized F layer. The AR is suppressed due
to the imbalance between up and down spin populations. Moreover,
the wave-vector mismatch between quasiparticles with up and down
spins in F leads to breaking of the retro-reflectivity of the AR
process.\cite{Kashiwaya} This means that after two consecutive
Andreev and normal reflections the quasiparticle will reach the
S/F interface at some distance, L, from the originally impinging
quasiparticle and thus a closed trajectory will not be realized.
However, as long as $L <\xi_{S}$, ABS can still be formed. The
above constraints, and the virtual Andreev reflection
process\cite{Kashiwaya, Zutic}, will result in a decreased
spectral weight of the ABS. Nevertheless, the penetration of ABS
into an F layer is still possible and is observed here by the
pronounced ZBCP we measured in the DOS of the F layer. The reduced
spectral weight is also demonstrated in the above data (figures 2
and 4). The ABS will survive in the F layer as long as its
thickness is smaller then the mean free path and phase coherence
is maintained between the electron-like and hole-like
quasiparticles along their trajectories. Our results show a clear
penetration of the ABS up to thicknesses of 9 nm. The mean free
path in our SRO islands could be significantly smaller than 50 nm,
the value measured in single-crystals at 1.8 K.\cite{Klein} Phase
coherence in F is maintained on the length-scale of the the
F-coherence length, which is estimated to be 3 nm in the clean
limit.\cite{Asulin2} Therefore, one should not expect penetration
to distances larger than $\sim$ 9 nm. We emphasize that the
measured ZBCP was continuous all over the SRO crystallites and was
not localized along narrow and elongated strips. For that reason,
the observed penetration of ABS into the SRO layer can not be
accounted for by effects related to magnetic domain walls, as was
the case in our previous work on (110)YBCO/SRO
(antinodal-YBCO/SRO) bilayers. \cite{Asulin2} In that study, the
OP penetrated the F layer to distances above 26 nm, an order of
magnitude larger than the F-coherence length. This penetration,
however,  took place only along narrow and elongated strips,
separated by at least 200 nm, consistent with the known magnetic
domain wall structure in $SrRuO_{3}$. This behavior was attributed
to crossed Andreev reflections, taking place in the vicinity of
the magnetic domain walls.\cite{Asulin2}\\
\indent The observed imbalance between peak heights is a
consequence of the spin polarization inside the F layer: the peak
that corresponds to the minority spins in F is suppressed with
respect to that corresponding to the majority spins. However, the
exchange field in F is not expected to split the penetrating ABS,
as it does not split the coherence peaks of the
observed\cite{Buzdin,Kontos} induced gaps in S. We therefore
suggest that the splitting of the ABS takes place at the interface
or inside S. Our observation that the negative and positive energy
peaks are suppressed with the same occurrence supports this
conjecture. When the spin degeneracy of the ABS is lifted and the
ZBCP is split, the peak that corresponds to the minority
(majority) spins is shifted to positive (negative) energies.
Consequently, if the splitting occurs inside F, then the spin
polarization should always suppress the peak that corresponds to
the minority spins, i. e., the suppression is always expected to
occur in the same side regardless of the direction of
magnetization in F. Which of the sides (positive or negative) will
be suppressed in F is determined by the sign of its spin
polarization, which is known to be negative in SRO (i.e., the
majority spin at the Fermi surface is antiparallel to the bulk
magnetization).\cite{Geballe} Therefore, only the negative peak is
expected to be suppressed in that case, in contrast to our
observation. This strongly indicates that the splitting cannot
occur inside the F islands. Moreover, the lack of correlation
between the degree of splitting and imbalance provides further
evidence that the splitting does not occur inside F. We believe
that the observed splitting is a consequence of an induced
magnetization inside S that penetrates it to a distance of
$\xi_{S}$. As noted above, this induced magnetization can be
parallel or antiparallel to that in F, depending on local changes
in the thickness of the F layer and the interface transparency.
\cite{Kharitonov,Bergeret1,Bergeret2,Krivoruchko} Therefore, both
negative and positive sides can be suppressed (on the same
sample), depending on the relative direction of
magnetization with respect to that of F.\\
\indent We now turn to discuss the magnitude of the split. The
relevant energy scale is the exchange energy, $E_{ex} \sim
K_{B}T_{Curie}$, which sets the upper limit for the value of the
energy shift. In SRO, $E_{ex}\sim 13 meV$ and would correspond to
a maximum energy shift of $\sim6.5$ meV to each side. The induced
magnetization should be weaker and hence the lower values we
typically observe. As demonstrated in figures 2 and 4, the
observed energy shift can change at different lateral locations on
the same F layer, even on the same crystallite. This can be
explained by local changes in thickness of the F layer and/or the
interface transparency that cause a non-uniformity in the induced
magnetization. In general, the energy shift (or split amplitude)
become larger as we move towards the center of the island, where
the thickness is usually larger compared to the rims.  This is
consistent with stronger magnetization and consequently stronger
induced magnetization in S, yielding larger splits (shifts) at the
YBCO side of the interface. The magnetic non-uniformity of the SRO
layer, manifested in the distribution of the SRO's $T_{Curie}$ in
Fig. 1b, supports this assumption. The junction transparency, on
the other hand, is not expected to change within an island over
such distances.  So, it seems that the local magnetization affects
our spectra, although the interface transparency and its
variations (mainly from island to
island) may also play an important role.\\
\indent A theoretical model that is seemingly related to our
experiment was introduced by Kashiwaya et al., who studied the
conductance properties of ferromagnet/ferromagnet-insulator/d-wave
superconductor junctions. \cite{Kashiwaya} Their calculated I-V
characteristics exhibit (in the case of the nodal S surface)
spectral features similar to some of those observed in our
measurements, namely, an asymmetric split of the ZBCP, where the
peak at positive energy is suppressed.  As in our case, the
polarization of the F electrode suppresses the peak that
corresponds to the minority spins. However, in their model the
splitting of the ZBCP is a consequence of 'spin filtering',
namely, different effective barrier heights that are felt by
quasi-particles with opposite spins that tunnel through the
ferromagnetic insulator. \cite{Kashiwaya} In spite of the fact
that we could achieve good fits to our spectra using this model,
we do not believe that it applies to our experiment, as explained
below. Very low F/S interface transparencies are needed in order
to reproduce our spectral features by this model, even lower than
the typical values corresponding to the tunnel junction between
the STM tip and the sample ($Z\sim5$ within the BTK formalism
\cite{BTK}).  Such low transparencies are not consistent, to say
the least, with our previous STS studies of SRO/YBCO bilayers
\cite{Asulin2} and magnetoresistance investigations in
YBCO/SRO/YBCO trilayers \cite {Aronov} where proximity effects
were observed, implying highly transparent (low $Z$ of less than
1) interfaces (that also allow for induced magnetization in the
YBCO layer). Consequently, in our measurements transport is
dominated by the tip-sample tunnel junction and therefore our
spectra reflect the DOS at the SRO surface and thus monitor the
penetration of the ABS into the SRO layer. In contrast, the I-V
spectra calculated by the Kashiwaya et al. model reflect the
properties of the S/F interface, in particular the effect of spin
filtering. We also like to point out that the spin filtering
scenario cannot naturally account, in our bilayers, for those
spectra in which the negative peak is suppressed. For this to
occur, one has to impose the existence of an insulating
ferromagnetic layer (between the YBCO and SRO layers) having spin
polarization opposite to that of the bulk SRO, which is
unrealistic.
\section{summary}
\indent In summary, our scanning tunneling spectroscopy
measurements of $SrRuO_{3}/(110)YBa_2Cu_3O_{7-\delta}$ bilayers
clearly reveal a penetration of the ABS into the F layer and
provide evidence for the predicted splitting of ABS at the S/F
interfaces due to an effect of induced magnetization at the S
side.
\section{Acknowledgments}
\indent We thank G. Deutscher, D. Orgad and L. Klein for helpful
discussions. This work was supported by the Israel Science
Foundation, Center of Excellence program, (grant \# 1565/04). O.
M. acknowledges the Harry deJur chair of applied sciences at the
Hebrew University.

\end{document}